\documentclass{Interspeech}

\interspeechcameraready

\title{DGMO: Training-Free Audio Source Separation through \\Diffusion-Guided Mask Optimization}

\author[affiliation={1}, equalcontribution]{Geonyoung}{Lee}
\author[affiliation={1}, equalcontribution]{Geonhee}{Han}
\author[affiliation={1}]{Paul Hongsuck}{Seo}

\affiliation{Department of Computer Science and Engineering}{Korea University}{Korea}
\email{gun0lee@korea.ac.kr, rtrt505@korea.ac.kr, phseo@korea.ac.kr}

\usepackage{comment}
\usepackage{multirow}
\usepackage{makecell}
\usepackage[normalem]{ulem}  
\usepackage{xcolor}          

\usepackage[capitalize]{cleveref}

\newcommand{\model}{DGMO\xspace}

\DeclareMathOperator*{\argmin}{argmin}

\begin{document}

\maketitle

\begin{abstract}
    Language-queried Audio Source Separation (LASS) enables open-vocabulary sound separation via natural language queries.
    While existing methods rely on task-specific training, we explore whether pretrained diffusion models, originally designed for audio generation, can inherently perform separation without further training. 
    In this study, we introduce a training-free framework leveraging generative priors for zero-shot LASS. 
    Analyzing na\"ive adaptations, we identify key limitations arising from modality-specific challenges.
    To address these issues, we propose Diffusion-Guided Mask Optimization (DGMO), a test-time optimization framework that refines spectrogram masks for precise, input-aligned separation.
    Our approach effectively repurposes pretrained diffusion models for source separation, achieving competitive performance without task-specific supervision. 
    This work expands the application of diffusion models beyond generation, establishing a new paradigm for zero-shot audio separation.{\footnote{The code is available at: \url{https://wltschmrz.github.io/DGMO/}.}} 

\end{abstract}

\begin{figure*}[t]
\vspace*{-1.55cm}
  \hspace*{-0.5cm}
  \centering
  \includegraphics[width=\textwidth]{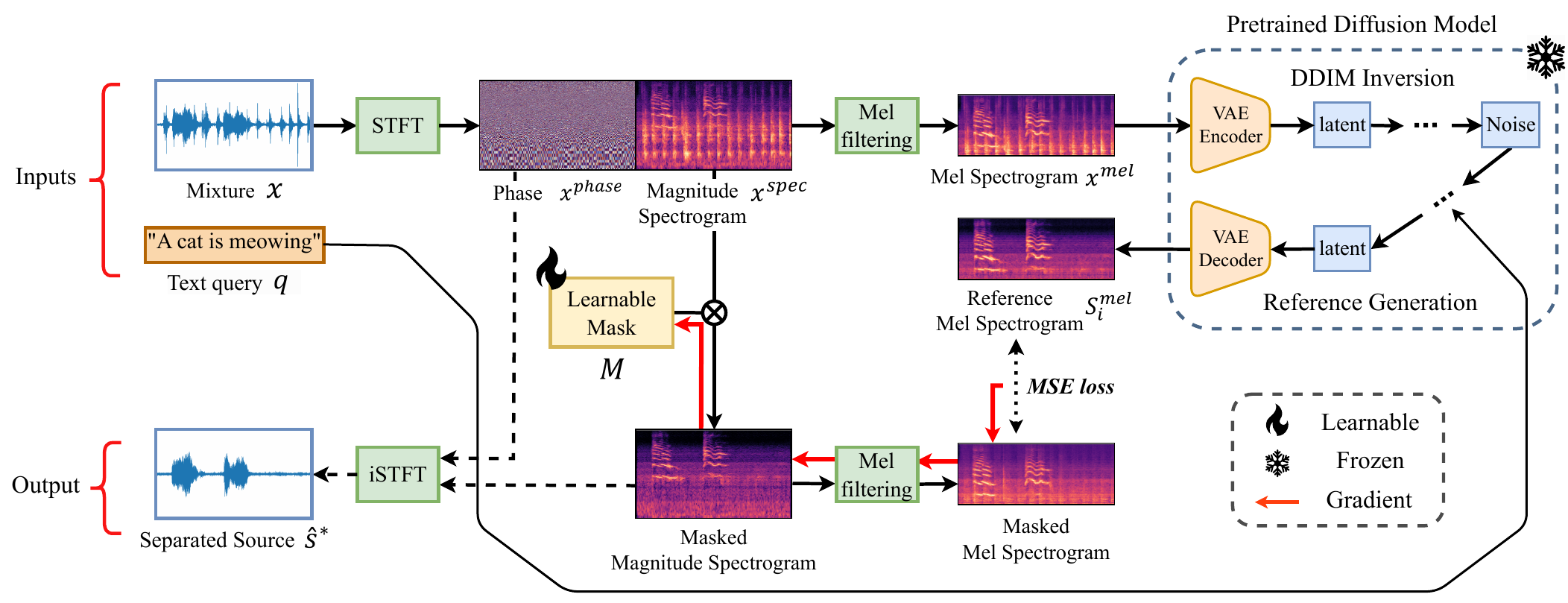}
  \caption{Training-free LASS framework using pre-trained diffusion model. It has two key processes: a Reference Generation and a Mask Optimization.}
  \label{fig:Total_pipeline_of_our_framework.}
\end{figure*}

\section{Introduction}

Humans can focus on specific sounds in complex auditory environments, a phenomenon known as the cocktail party effect\cite{haykin2005cocktail}.
Computational models aim to replicate this ability through sound separation, isolating target sources from audio mixtures. 
Language-queried Audio Source Separation (LASS) has emerged as a flexible solution, allowing users to specify target sounds via natural language queries~\cite{lassnet2022, clipsep2022, audiosep2024}.
However, existing LASS models predominantly rely on task-specific training, where networks are explicitly trained for sound separation. 
Recent advances have explored generative models for LASS~\cite{flowsep2024, soloaudio2025}, but these methods still require specialized training, limiting their flexibility and scalability across different domains.

In this study, we introduce a training-free framework that repurposes pretrained generative models for source separation. 
Diffusion models, which have demonstrated remarkable performance in audio generation~\cite{audioldm2023, auffusion2024}, remain largely unexplored for sound separation.
Unlike prior LASS methods that require task-specific training, we investigate whether a pretrained generative model can inherently perform separation without further training for this task.
Our approach leverages diffusion models’ generalization ability, enabling zero-shot separation by extracting sound sources based on textual queries.
To explore diffusion-based LASS, we first investigate na\"ive adaptations, such as input mask optimization—an approach previously used in referring image segmentation~\cite{peekaboo2022}, which is conceptually related to source separation.
However, applying diffusion models to audio separation presents unique challenges due to the fundamental differences between audio and visual modalities including phase inconsistencies and the need for precise time alignment.
To overcome these challenges, we propose Diffusion-Guided Mask Optimization (DGMO), a test-time framework that integrates generative priors with explicit mask opimization. 
Rather than treating separation as a purely generative process, DGMO refines a learnable mask in the magnitude spectrogram domain, ensuring time alignment while leveraging diffusion-generated references in the mel spectrogram domain.
This hybrid approach preserves the fidelity of separated audio, mitigating artifacts and inconsistencies seen in previous na\"ive generative methods~\cite{flowsep2024, soloaudio2025}.

Our key contributions are as follows: (1) We establish a fully training-free framework by repurposing diffusion models for audio separation without additional training. (2) We identify limitations in naïve adaptations of diffusion models to LASS and propose Diffusion-Guided Mask Optimization (DGMO), a test-time optimization framework overcoming the unique challenges in the audio modality. (3) To the best of our knowledge, this is the first work to apply pretrained generative models to training-free, zero-shot source separation, expanding the role of diffusion models beyond generation.

\section{Related Works}

\noindent \textbf{Language-queried Audio Source Separation} \ \ Early sound separation models achieved success within predefined domains~\cite{music_seperation, sppech_seperation, audio_event_seperation}. Research has since expanded to universal sound sources using vision~\cite{sop2018}, audio~\cite{uss_weak_data}, label~\cite{clipsep2022}, and language queries. The language-based approach is appealing for its accessibility. LASS-Net~\cite{lassnet2022} first introduced a BERT-based text encoder but required joint text-audio optimization. With multimodal learning advancements~\cite{clap2022, clap2023, clip}, methods aligning modalities in a shared space emerged, reducing alignment constraints~\cite{clipsep2022, audiosep2024}. Moreover, generative approaches for LASS~\cite{flowsep2024, soloaudio2025} have been proposed to directly synthesize the target audio.

\noindent \textbf{Diffusion Models and Non-Generation Tasks} \ \ Diffusion models excel in text-to-image~\cite{stable_diffusion, mmdit} and text-to-audio tasks. AudioLDM~\cite{audioldm2023, audioldm2_2024} and Auffusion~\cite{auffusion2024} leverage latent diffusion for realistic audio synthesis. Beyond generation, they enhance test-time optimization and editing. DreamFusion~\cite{dreamfusion2022} applies score distillation sampling for 3D synthesis, while Peekaboo~\cite{peekaboo2022} refines segmentation via inference-time mask optimization. Furthermore, audio editing methods~\cite{audio_edit, audio_prompt_edit} and image inversion techniques~\cite{ddim, null2022} demonstrate diffusion models' versatility in refining and manipulating signals.

\section{Method}

\subsection{Language-queried Audio Source Separation}

Given an audio mixture $x$ composed of multiple source signals $\{s_i\}$ and environmental noise $e$ formulated as $x = \sum_i s_i + e$, LASS~\cite{lassnet2022} aims to extract a target source $s^*$ described by a natural language query $q$.
Conventionally, this task is addressed by estimating a mask $M(x, q)$ and applying it to the mixture, such that $s^*=x \odot M$, where $\odot$ denotes the element-wise multiplication, preventing additional artifacts that may arise from directly generating signals.
By leveraging textual descriptions instead of predefined categories, LASS enables flexible and intuitive audio separation. 
However, this task requires learning cross-modal associations between natural language queries and audio sources, posing significant challenges in achieving precise text-audio alignment.
This challenge has led prior work to train task-specific models for learning such associations.
In contrast, we explore the capability of pretrained diffusion models~\cite{audioldm2023, auffusion2024, audioldm2_2024} originally designed for audio generation, to perform source separation without any task-specific training,
leveraging their inherent generative priors for zero-shot language-queried audio source separation. 

\subsection{Diffusion Models and Mask Optimization}
\label{sec:mask_optim}

Diffusion models~\cite{ddpm} are generative models that iteratively refine noisy inputs by learning a data distribution through a forward noise-injection and reverse denoising process. 
The reverse process estimates the original data by predicting and removing noise, formulated as:
\begin{equation}
    x_{t-1} = \frac{1}{\sqrt{\alpha_t}} 
    \left( x_t - \frac{1 - \alpha_t}{\sqrt{1-\bar{\alpha}_t}}{\epsilon}_\theta(x_t, q, t) \right) + \sigma_t z
\end{equation}
where $\alpha_t$ is the noise scaling factor, $\epsilon_\theta(x_t,q,t)$ is the predicted noise conditioned on the noisy input $x_t$, language query $q$, and timestep $t$, and $z\sim\mathcal{N}(0,I)$ is a standard Gaussian noise term with $\sigma_t$ controlling the variance of the stochastic update.

While diffusion models have demonstrated high-quality generation including in the audio domain, their potential for signal separation remains largely unexplored. 
A notable exception is \cite{peekaboo2022}, a prior approach in computer vision that performs test-time optimization using score distillation loss with a pretrained diffusion model for segmentation based on a language query—an approach analogous to sound separation, as both tasks aim to isolate distinct components from an input mixture.
Specifically, $x_0$ is masked by $M$ before the noise injection and $M$ is optimized to minimize the diffusion loss function\footnote{For notational simplicity, we present equations using regular diffusion models, though our experiments utilize latent diffusion models.}: 
\begin{align}
    x_t &= \sqrt{\bar{\alpha}_t} (x_0 \odot M) + \sqrt{1 - \bar{\alpha}_t} \epsilon \\
    M^* &= \argmin_{M} \mathbb{E}_{\epsilon, t} 
    \Big[
    w_t \cdot \Vert \hat{\epsilon}_{\theta} (x_t, q, t) - \epsilon \Vert_2^2
    \Big]
\end{align}
where $w_t$ is a weighting term computed from noise schedule parameters that depends on timestep $t$.
Through this optimization process, the optimal mask $M^*$ learns to remove irrelevant regions of the input image $x_0$, ensuring it best corresponds to the query $q$ effectively achieving segmentation.

Given the similarity between image segmentation and LASS, one may think that we can directly apply above technique to LASS.
However, unlike visual signals, which are non-additive due to occlusion—where objects can block and completely remove parts of other objects—audio signals are additive, meaning multiple sources mix without fully masking each other. 
Therefore, to separate audio signals through masking, we cannot simply apply a binary mask as in visual segmentation. 
Unlike in the visual domain, where occluded parts can be directly masked out, audio separation requires computing the remaining audio signals to be removed, making the process as challenging as directly generating the target sound. 
This poses a unique challenge in the audio domain, preventing the above mask optimization with diffusion models from succeeding in the same way it does for visual segmentation.

\subsection{Separated Audio Generation}
\label{sec:sep_audio}
An alternative approach to constructing separated audio signals using a diffusion model is to generate the target sound $s^*$ directly from the query $q$, conditioned on the input mixture $x$.
This approach is inspired by inversion-based editing techniques~\cite{null2022}, where a model refines an existing signal to align with a given target representation by $q$.

Specifically, a denoised output $x_0$ can be generated from a noised input $x_t$, which is derived from the original mixture $x$. 
With an appropriately chosen $t$, the reconstructed $x_0$ serves as the separated audio, as it retains the essential content semantics of $x$ while being regenerated under the condition $q$, effectively filtering out mismatching components.
In this process, the choice of $t$ is crucial: if too large, $x_t$ may lose essential attributes from $x$, while if too small, it may not introduce enough noise for effective regeneration.
A well-balanced $t$ ensures that relevant information is preserved while allowing the model to refine the signal to align with the given query.

While this regeneration technique effectively generates sounds relevant to the query $q$ and resembles the original source within $x$, the generated outputs often introduce artifacts or contain entirely new sounds that only superficially match the intended target, lacking true correspondence to the original signal.
This highlights the need for an explicit constraint, similar to the mask optimization process, to ensure that the generated output remains faithful to the original source while effectively isolating the target sound.

\subsection{Diffusion-Guided Mask Optimization}
We propose a novel training-free LASS framework based on diffusion models, which overcomes the limitations of previous approaches by integrating both mask optimization and generative refinement into a unified process.
This framework operates in two stages: reference generation and mask optimization.

\noindent \textbf{Reference Generation} \ \ 
In this stage, we generate separated audio given $x$ and $q$ following the procedure in \cref{sec:sep_audio}, referring to the generated audio signals $\{s_i\}$ as references.
As discussed, these references inherit attributes from $x$ but often introduce sound elements that are not originally present in $x$ due to the absence of explicit constraints, which are difficult to impose effectively within a diffusion model.
 
\noindent \textbf{Mask Optimization} \ \ 
Once the reference signals $\{s_i\}$ are generated, they encapsulate the knowledge embedded within the diffusion model regarding both the input mixture $x$ and the query $q$.
However, since there is no explicit constraint that ensures the separated sound $s^*$ strictly belongs to the mixture $x$, we introduce a mask optimization process to enforce consistency with the input mixture.
Specifically, rather than using the references directly as separated outputs, we use them as supervision signals to guide a mask $M$ applied to the mixture $x$.

Since diffusion models operate in the mel spectrogram domain, we define the optimization loss by comparing the mel spectrograms of the masked mixture and the reference signal.
However, applying the mask directly in the mel domain is infeasible due to the lossy, non-invertible mel transformation, which prohibits faithful waveform reconstruction.
While vocoder-based reconstruction can be used to directly convert mel spectrograms back to waveforms, it typically induces temporal artifacts and alignment errors, as it generates phase through neural prediction instead of retaining the mixture’s true phase.

To mitigate these issues, we decouple the optimization and evaluation spaces: the mask is applied in the magnitude spectrogram domain for stable and interpretable reconstruction, while the loss is computed in the mel domain to maintain compatibility with the model’s conditioning. Formally, for each reference $s_i$, we define the objective as:
\begin{equation}
    \mathcal{L}_i(M) = \Vert \mathrm{mel}(x^\mathrm{spec} \odot M) - s_i^\mathrm{mel} \Vert^2_2
\end{equation}
where $x^\mathrm{spec}$ is the magnitude spectrogram of the input mixture, $M$ is the mask, and $s_i^{\mathrm{mel}}$ is the mel-spectrogram of the corresponding reference $s_i$. This formulation enables effective gradient-based optimization while ensuring the extracted output remains both physically plausible and semantically aligned.

To improve robustness, we average the individual losses with multiple references $\{s_1, \dots, s_n\}$:
\begin{equation}
    M^* = \argmin_M \frac{1}{n} \sum_{i=1}^n \mathcal{L}_i(M)
\end{equation}
Using multiple references mitigates high variance in mask optimization, as each reference captures different aspects of the target source. All components in this process are differentiable, allowing gradient-based optimization.

The estimated target waveform $\hat{s}^*$ is then reconstructed using the optimized mask $M^*$ and the original phase:
\begin{equation}
    \hat{s}^* = \mathrm{iSTFT}(x^\mathrm{phase}, x^\mathrm{spec} \odot M^*)
\end{equation}
Here, $x^\mathrm{spec}$ and $x^\mathrm{phase}$ denote the magnitude and phase spectrograms of the mixture $x$, respectively.

\noindent \textbf{DDIM Inversion} \ \ 
A na\"ive approach for reference generation injects random Gaussian noise into the input mixture. 
However, such arbitrary noise overwrites the structure and source-related signals, resulting in outputs that deviate from the original mixture content. 
While reducing the noise level might help retain more structure, it hampers the removal of non-target components. 
To address this, we adopt DDIM inversion~\cite{ddim, null2022}, a deterministic alternative that transforms the input mixture $x_0$ into a noisy $x_t$ without randomness.
Unlike random noise injection, DDIM inversion preserves the content structure of $x_0$ and maintains semantic fidelity throughout the reference generation process. This improvement ensures reliable reference signals, facilitating effective mask optimization.

\begin{table}[t]
\vspace*{-1.32cm}
  \caption{\textbf{Evaluation of Diffusion based LASS approaches.} We present results on the AudioCaps dataset. Numbers reported are SI-SDR and SDRi values for our method and baselines.}
  \label{tab:ours_vs_baselines}
  \centering
  \scalebox{0.9}{
  \begin{tabular}{l | r  r}
    \toprule
    \multicolumn{1}{c}{\textbf{Models}} & 
        \multicolumn{1}{r}{\textbf{SI-SDR}} &
            \multicolumn{1}{r}{\textbf{SDRi}} \\
    \midrule
    Original Mixture (No Separation) & $-0.07$ & $0$ \\
    Mask Optimization (\cref{sec:mask_optim}) & $-0.06$ & $2.33$ \\
    Separated Audio Generation (\cref{sec:sep_audio}) & $0.20$ & $-0.24$ \\
    \midrule
    Diffusion-Guided Mask Optimization (Ours) & $\textbf{1.99}$ & $\textbf{3.57}$ \\
    \bottomrule
  \end{tabular}
  }
\end{table}

\begin{table*}[t]
\vspace*{-1.3cm}
  \caption{Benchmark evaluation results of DGMO and comparison with state-of-the-art LASS systems. For CLAP scores, except for our model, the results are sourced from ~\cite{flowsep2024}.}
  \label{tab:benchmark_results}
  \centering
  \scalebox{0.78}{
  \begin{tabular}{ll|ccc|ccc|ccc|ccc}
    \toprule
    \multicolumn{2}{c|}{} & 
        \multicolumn{3}{c}{\textbf{VGGSound}} &
        \multicolumn{3}{c}{\textbf{AudioCaps}} &
        \multicolumn{3}{c}{\textbf{MUSIC}} &
        \multicolumn{3}{c}{\textbf{ESC-50}}  \\
    \cmidrule(lr){3-14}
    \textbf{Training Type} & \textbf{Models} & 
    SI-SDR & SDRi & \text{CLAP}$_\text{Score}$ & 
    SI-SDR & SDRi & \text{CLAP}$_\text{Score}$ & 
    SI-SDR & SDRi & \text{CLAP}$_\text{Score}$ &
    SI-SDR & SDRi & \text{CLAP}$_\text{Score}$ \\
    \midrule
    \multirow{3}{*}{\makecell[l]{Supervised \\ training}}
    & LASSNet \cite{lassnet2022} &
    -4.50 & 1.17 & 17.40 &
    -0.96 & 3.32 & 14.40 &
    -13.55 & 0.13 & - &
    -2.11 & 3.69 & 20.50 \\
    
    & CLIPSep \cite{clipsep2022}  &
    1.22 & 3.18 & - &
    -0.09 & 2.95 & - &
    -0.37 & 2.50 & - &
    -0.68 & 2.64 & - \\
    
    & AudioSep \cite{audiosep2024} &
    9.04 & 9.14 &  19.00 &
    7.19 & 8.22 & 13.60 &
    9.43 & 10.51 & - &
    8.81 & 10.04 & 21.20 \\

    \midrule
    Train-free & Ours &
    1.80 & 2.65 & 18.70 &
    1.89 & 3.62 & 18.60 &
    0.56 & 2.82 & 24.60 &
    1.98 & 3.27 & 22.00  \\
    \bottomrule
  \end{tabular}
}
\end{table*}
\section{Experiments}

\subsection{Evaluation Benchmarks}
For evaluation, we use four publicly available text-aligned audio datasets and construct artificial mixtures following prior research in LASS~\cite{clipsep2022, audiosep2024}.
All datasets include both training and test sets. 
However, as our method is entirely training-free, we exclusively utilize the test set for evaluation.
separation models.

\noindent \textbf{VGGSound}~\cite{vggsound2020} \ \ 
We adopt the evaluation setup of~\cite{audiosep2024}, where 100 clean target audio samples are each mixed with 10 randomly selected background samples from the test set. Loudness is uniformly sampled between -35 dB and -25 dB LUFS, and mixtures are normalized to 0.9 if clipping occurs, resulting in 1,000 mixtures with an average SNR of ~0 dB.

\noindent \textbf{AudioCaps}~\cite{audiocaps2019} \ \ 
We follow~\cite{audiosep2024}, where the AudioCaps test set of 957 audio clips, each with five captions, is used to construct 4,785 mixtures for LASS.
Each target source is mixed with five randomly selected background sources with different sound event tags.
Mixtures are generated at 0 dB SNR, ensuring equal energy levels between the target and background sounds.

\noindent \textbf{MUSIC}~\cite{sop2018} \ \ 
MUSIC contains 536 high-quality videos of 11 musical instruments sourced from YouTube.
Following~\cite{clipsep2022}, 5,004 test examples for sound source separation constructed from 46 test videos from MUSIC by mixing randomly selected segments from different instrument classes at an SNR of 0 dB.

\noindent \textbf{ESC-50}~\cite{esc2015} \ \ 
While the dataset contains 2,000 audio clips across 50 classes, mixtures are created by pairing clips from different classes at 0 dB SNR. 
Constructing 40 mixtures per class, it contains 2,000 evaluation pairs.
\begin{table}[t]
\vspace*{-0.2cm}
  \caption{\textbf{\model with Various Diffusion Models.} It presents the performance of \model applied to different models. Results are evaluated on the AudioCaps test set with 100 samples. Metrics reported are SI-SDR and SDRi. Additionally, we present FAD of these models, which are taken from the~\cite{auffusion2024}, where lower values indicate better generation performance by measuring the distance between generated and real audio distributions.}
  \label{tab:tta_model_comparison}
  \centering
  \scalebox{0.86}{
  \begin{tabular}{l | c | cc}
    \toprule
    \textbf{Audio Diffusion Model} & 
    \textbf{FAD (Generation)} &
    \textbf{SI-SDR} & 
    \textbf{SDRi} \\
    \midrule
    AudioLDM~\cite{audioldm2023}  & $4.40$ & $1.10$ & $3.12$  \\
    AudioLDM2~\cite{audioldm2_2024} & $2.19$ & $1.58$ & $2.89$ \\
    Auffusion~\cite{auffusion2024} & $1.63$ & $\textbf{1.99}$ & $\textbf{3.57}$  \\
    \bottomrule
  \end{tabular}
  }
\end{table}

\subsection{Evaluation Metrics} 
We evaluate the performance of our methods using three widely adopted metrics: scale-invariant source-to-distortion ratio~\cite{sisdr2019} (SI-SDR), signal-to-distortion ratio improvement~\cite{uss_weak_data} (SDRi), and CLAP Score~\cite{clap_score}. SI-SDR measures the quality of separated signals by assessing residual distortion and interference, independent of signal scale. SDRi quantifies the improvement in separation quality relative to the original mixture, providing a comparative measure of enhancement. CLAP Score, a reference-free metric, evaluates the semantic alignment between the separated audio and the text prompt, reflecting how well the output matches the intended content. Higher values across all metrics indicate better separation performance.

\subsection{Implementation Details}
We use the pre-trained text-to-audio diffusion model, Auffusion~\cite{auffusion2024}, following the original diffusion model’s preprocessing. Audio is sampled at 16\,kHz, padded to 10.24\,s, then centered and normalized. We apply STFT with 256 mel filter banks, a window length of 1024, an FFT size of 2048, and a hop length of 160. For reference generation, DDIM inversion is performed in 25 steps with a noising step ratio of 0.7 and null text. We sample references in batches of 4 and optimize masks for 300 epochs per iteration, over 2 iterations.

\subsection{Results}
\noindent \textbf{Comparisons to Na\"ive Approaches} \ \ 
\cref{tab:ours_vs_baselines} compares the proposed method with the na\"ive approaches described in Sections~\ref{sec:mask_optim} and \ref{sec:sep_audio}.
The na\"ive mask optimization method completely fails to find separation masks resulting in even lower scores than the original mixture $x$ due to the complexity of the task.
The separated audio generation technique improves scores but its effectiveness is limited, as the generated audio often contains signals not originally present in $x$.
In contrast, the proposed diffusion-guided mask optimization successfully separates the target sound using only a pretrained diffusion model without any task-specific training.

\noindent \textbf{Comparisons to Supervised Methods}
We compare our method with other supervised methods.
LASS-Net~\cite{lassnet2022} uses a pre-trained BERT~\cite{devlin2018bert} and ResUNet~\cite{diakogiannis2020resunet}.
CLIPSep~\cite{clipsep2022} employs CLIP~\cite{clip} and SOP~\cite{sop2018}.
Both models operate in the frequency domain and reconstruct waveforms using noisy phase information. 
AudioSep~\cite{audiosep2024} also employs the CLAP and trained with captioning data~\cite{audiocaps2019, clotho2020}.
We report the evaluation results as provided in prior work~\cite{flowsep2024, clap_score}, where models were assessed on the same dataset using predefined metrics

\noindent \textbf{Ablations with Various Diffusion Models} \ \ 
We evaluate our framework using multiple audio diffusion models.
As shown in \cref{tab:tta_model_comparison}, our framework performs consistently well across different models, demonstrating its robustness. 
Additionally, the zero-shot separation performance generally aligns with the audio generation quality of each model (\textit{e.g.}, SI-SDR vs. FAD for generation), indicating a strong correlation between a model's generative capability and its effectiveness in source separation.

\noindent \textbf{Effects of DDIM Inversion and Noising Step $t$} \ \ 
\cref{tab:ddim_inversion_ratio} shows performance variations across different noising steps $t$. With random noise injection, too small a ratio $t/T$ introduces insufficient noise, degrading separation quality. As the ratio increases, the injected noise dominates, reducing the correlation between the original input and the resulting signal. In contrast, DDIM inversion shows stable and superior performance across all noise scales. By leveraging structured, content-aware noise injection, it consistently mitigates the trade-off observed in random noise injection. These results highlight the robustness and effectiveness of DDIM inversion across different noise scales, reinforcing its suitability for source separation tasks.

\begin{table}[t]
\vspace*{-0.2cm}
\caption{\textbf{Effect of Noising Step Ratio on \model Performance.} Performance of \model with varying noising step ratios, evaluated on the AudioCaps test set (100 samples). The results demonstrate how the inversion ratio influences audio source separation quality. Metrics reported are SI-SDR and SDRi.}
\label{tab:ddim_inversion_ratio}
\centering
\scalebox{0.84}{ 
\begin{tabular}{l | l | c c c c c}
\toprule
& & \multicolumn{5}{c}{\textbf{Noising Step Ratio (t/T)}} \\
\cmidrule(lr){3-7}
\textbf{Method} & \textbf{Metric} & \textbf{0.1} & \textbf{0.3} & \textbf{0.5} & \textbf{0.7} & \textbf{0.9} \\
\midrule
Random & SI-SDR & -0.59 & -0.79 & -0.80 & -0.86 & -1.05 \\
Noise Injection& SDRi & 2.28 & 2.57 & 2.68 & 1.60 & 2.48 \\
\midrule
\multirow{2}{*}{DDIM Inversion} & SI-SDR & -0.57 & -0.34 & 0.62 & 1.99 & 2.04 \\
& SDRi & 2.71 & 2.81 & 3.15 & 3.57 & 3.64 \\
\bottomrule
\end{tabular}
}
\end{table}

\section{Conclusion}

We explored the feasibility of training-free LASS by leveraging pretrained diffusion models, originally designed for audio generation, for zero-shot source separation. 
We analyzed na\"ive adaptations of diffusion models to LASS and identified key limitations. 
To address these challenges, we introduced Diffusion-Guided Mask Optimization, a test-time optimization framework that refines spectrogram masks for accurate, input-aligned separation.
Our results demonstrate that pretrained generative models can be effectively repurposed for source separation without task-specific training, achieving competitive performance.

\section{Acknowledgements}
This research was supported by IITP grants
(IITP\allowbreak-2025-RS-2020-II201819,
IITP-2025-RS-2024-00436857, 
IITP-2025-RS-2024-00398115,
IITP-2025-RS-2025-02263754, 
IITP-2025-RS-\allowbreak2025-02304828
), and the KOCCA grant (RS-2024-00345025
) funded by the Korea government (MSIT, MOE and MSCT).

\bibliographystyle{IEEEtran}
\bibliography{mybib}

\end{document}